\documentclass{SciPost}

\hypersetup{
    colorlinks,
    linkcolor={red!50!black},
    citecolor={blue!50!black},
    urlcolor={blue!80!black}
}

\usepackage[bitstream-charter]{mathdesign}
\DeclareSymbolFont{usualmathcal}{OMS}{cmsy}{m}{n}
\DeclareSymbolFontAlphabet{\mathcal}{usualmathcal}

\fancypagestyle{SPstyle}{
\fancyhf{}
\lhead{\colorbox{scipostblue}{\bf \color{white} ~SciPost Physics }}
\rhead{{\bf \color{scipostdeepblue} ~Submission }}

\fancyfoot[C]{\textbf{\thepage}}
}

\newcommand{\cor}[1]{#1}
\newcommand{\eg}{e.g.,~}
\newcommand{\ie}{i.e.,~}

\begin{document}

\pagestyle{SPstyle}

\begin{center}{\Large \textbf{\color{scipostdeepblue}{
%%%%%%%%%% TODO: Write your article's title here
On the maximum compactness of neutron stars\\
%%%%%%%%%% END TODO: TITLE
}}}\end{center}

\begin{center}\textbf{
%%%%%%%%%% TODO: AUTHORS
% Write the author list here. 
% Use (full) first name (+ middle name initials) + surname format.
% Separate subsequent authors by a comma, omit comma and use "and" for the last author.
% Mark the corresponding author(s) with a superscript symbol in this order
% \star, \dagger, \ddagger, \circ, \S, \P, \parallel, ...
Luciano Rezzolla\textsuperscript{1,2,3},
Christian Ecker\textsuperscript{1}
%%%%%%%%%% END TODO: AUTHORS
}\end{center}

\begin{center}
%%%%%%%%%% TODO: AFFILIATIONS
% Write all affiliations here.
% Format: institute, city, country
{\bf 1} Institut f\"ur Theoretische Physik, Goethe Universit\"at,
Max-von-Laue-Str. 1, D-60438 Frankfurt am Main, Germany\\
{\bf 2} CERN, Theoretical Physics Department, 1211 Geneva 23,
Switzerland\\
{\bf 3} School of Mathematics, Trinity College, Dublin, Ireland
%%%%%%%%%% END TODO: AFFILIATIONS
%% %%%%%%%%%% TODO: EMAIL
%% % Provide email address of corresponding author(s)
%% \\[\baselineskip]
%% $\star$ \href{mailto:email1}{\small email1}\,,\quad
%% $\dagger$ \href{mailto:email2}{\small email2}
%% %%%%%%%%%% END TODO: EMAIL
\end{center}

\section*{\color{scipostdeepblue}{Abstract}}
\textbf{\boldmath{%
%%%%%%%%%% TODO: ABSTRACT
% Write your abstract here.
%% The abstract is in boldface, and should fit in 8 lines. It should be
%% written in a clear and accessible style, emphasizing the context, the
%% problem(s) studied, the methods used, the results obtained, the
%% conclusions reached, and the outlook. You can add a table contents,
%% recommended if your paper is more than 6 pages long.
The stellar compactness, that is, the dimensionless ratio between the
mass and radius of a compact star, $\mathcal{C} := M/R$, plays a
fundamental role in characterising the gravitational and nuclear-physics
aspects of neutron stars. Yet, because the compactness depends
sensitively on the unknown equation of state (EOS) of nuclear matter, the
simple question: \textit{``how compact can a neutron star be?''} remains
unanswered. To address this question, we adopt a statistical approach and
consider a large number of parameterised EOSs that satisfy all known
constraints from nuclear theory, perturbative Quantum Chromodynamics
(QCD), and astrophysical observations. Next, we conjecture that, for any
given EOS, the maximum compactness is attained by the star with the
maximum mass of the sequence of nonrotating configurations. While we can
prove this conjecture for a rather large class of solutions, its general
proof is still lacking. However, the evidence from all of the EOSs
considered strongly indicates that it is true in general. Exploiting the
conjecture, we can concentrate on the compactness of the maximum-mass
stars and show that an upper limit appears for the maximum compactness
and is given by $\mathcal{C}_{\rm max} = 1/3$. Importantly, this upper
limit is essentially independent of the stellar mass and a direct
consequence of perturbative-QCD constraints.
%%%%%%%%%% END TODO: ABSTRACT
}}

\vspace{\baselineskip}

%% %%%%%%%%%% BLOCK: Copyright information
%% % This block will be filled during the proof stage, and finilized just before publication.
%% % It exists here only as a placeholder, and should not be modified by authors.
%% \noindent\textcolor{white!90!black}{%
%% \fbox{\parbox{0.975\linewidth}{%
%% \textcolor{white!40!black}{\begin{tabular}{lr}%
%%   \begin{minipage}{0.6\textwidth}%
%%     {\small Copyright attribution to authors. \newline
%%     This work is a submission to SciPost Physics. \newline
%%     License information to appear upon publication. \newline
%%     Publication information to appear upon publication.}
%%   \end{minipage} & \begin{minipage}{0.4\textwidth}
%%     {\small Received Date \newline Accepted Date \newline Published Date}%
%%   \end{minipage}
%% \end{tabular}}
%% }}
%% }
%% %%%%%%%%%% BLOCK: Copyright information

%%%%%%%%%% TODO: LINENO
% For convenience during refereeing we turn on line numbers:
%\linenumbers
% You should run LaTeX twice in order for the line numbers to appear.
%%%%%%%%%% END TODO: LINENO

%%%%%%%%%% TODO: TOC 
% Guideline: if your paper is longer that 6 pages, include a TOC
% To remove the TOC, simply cut the following block
\vspace{10pt}
\noindent\rule{\textwidth}{1pt}
\tableofcontents
\noindent\rule{\textwidth}{1pt}
\vspace{10pt}
%%%%%%%%%% END TODO: TOC

%%%%%%%%% TODO: CONTENTS 
% Write your article contents here, starting from first \section.
% An example structure is given below.

%========================================================================
\section{Introduction}
%========================================================================

Neutron stars are prime examples of extremely
compact astrophysical objects, where the physical conditions are so
extreme that all four fundamental forces of nature play a significant
role in determining their structure and dynamics~(see, \eg
\cite{Rezzolla2018} for a comprehensive collection). When considering
neutron stars as static and spherically symmetric solutions of the
Einstein equations for a self-gravitating fluid, two quantities are
particularly important: the mass $M$ and the radius $R$. While the masses
can be measured to very high precision thanks to accurate radio-pulsar
measurements~(see, \eg \cite{Antoniadis:2013pzd, NANOGrav:2019jur,
  Fonseca:2021wxt}), the radii are known only poorly, mostly because of
the complex physics that accompanies their surface emission~(see, \eg
\cite{Riley2019, MCMiller2019b}). The challenges associated with
performing a measurement of their size, combined with the enormous
theoretical challenges in describing the equation of state (EOS) that
regulates their structure and composition, make neutron stars as
fascinating as puzzling.

Once an EOS is prescribed, the solution of the Einstein equations for a
self-gravitating nonrotating fluid provides an infinite family of
equilibrium models characterised by specific values of the mass and
radius. A fundamental difference with respect to the equivalent family in
Newtonian gravity is that the relativistic (gravitational) mass is upper
bounded by a maximum value, $M_{_{\rm TOV}}$~(see, \eg \cite{MTW1973}),
where the index ``TOV'' refers to the Tolman-Oppenheimer-Volkoff
equations, whose solution is needed to obtain the equilibria~(see, \eg
\cite{Rezzolla_book:2013}). Put differently, while equilibrium solutions
can be constructed with central energy densities exceeding those
corresponding to $M_{_{\rm TOV}}$, these configurations yield $M <
M_{_{\rm TOV}}$ and are located on the unstable branch of the $M$-$R$
sequence.

Given the mass and radius of a neutron star, a derived quantity that
naturally appears in the properties describing the corresponding
spacetime is the compactness defined as 
\begin{equation}
  \mathcal{C} := \frac{G\,M}{c^2\,R}\,,
\end{equation}
where $G$ and $c$ are the gravitational constant and speed of light,
respectively. Hereafter, we will adopt geometric units in which $G=1=c$,
so that the compactness $\mathcal{C} = M/R$ is a dimensionless quantity.

It is well-known in general relativity that the compactness of a static,
spherically symmetric, vacuum spacetime is upper bounded by $\mathcal{C}
\leq 1/2$, where the equality refers to a black hole described by the
Schwarzschild solution. Also well-known in general relativity as the
``Buchdahl limit''~\cite{Buchdahl:59} is that the compactness of a
static, spherically symmetric, non-vacuum spacetime is upper bounded by
$\mathcal{C} \leq 4/9 =: \mathcal{C}_{\rm Buch}$~(see
also~\cite{Dadhich2017} for more general spacetimes). Finally, it is also
well-known that when employing microphysical EOSs models to describe the
nuclear matter composing neutron stars, the compactness reached are
generally smaller than those constrained by the Buchdahl bound, i.e.,
$\mathcal{C} \sim 0.1-0.2$ (\cor{see~\cite{Lindblom1984} for one of the
  early studies setting upper limits on the compactness}). At the same
time,~Haensel and Zdunik~\cite{Haensel1989} have argued that the most
compact configurations are produced when the low-density part of the EOS
is maximally soft, while the high-density is maximally stiff (\ie with
sound speed $c_s=1$), thus obtaining a maximum compactness of
$\mathcal{C}=0.3543 = 1/2.8294$~\cite{Koranda1997,
  Lattimer2016}. \cor{This bound is also known as the ``causality
  bound''.}

The purpose of this work is twofold. First, we propose a conjecture
according to which the maximum compactness is achieved by the star with
maximum mass, \ie $M_{_{\rm TOV}}$. In other words, we conjecture that
the star with maximum mass is also the star with the maximum compactness:
$\mathcal{C}_{\rm max} = \mathcal{C}_{_{\rm TOV}}:= M_{_{\rm
    TOV}}/R_{_{\rm TOV}}$. This conjecture can be shown to be
mathematically true for some specific non-vacuum spacetimes, for a large
class of generic spacetimes, and we provide numerical evidence that it
holds for all of the stellar models considered here. Second, we show that
when considering EOSs that satisfy constraints derived from nuclear
physics and astrophysics, the compactness of \textit{realistic} neutron
stars is also upper bounded $\mathcal{C} \leq \mathcal{C}_{\rm max} <
\mathcal{C}_{\rm Buch}$ and we further determine the value of
$\mathcal{C}_{\rm max}$.

In order to achieve our goals, we will employ a very large set of
parameterised EOSs constructed agnostically but such that they satisfy
all known constraints derived either from nuclear physics, from
perturbative Quantum Chromodynamics (pQCD), gravitational-wave detections,
or astrophysical observations of isolated neutron stars. In this way, we
show that for all the stellar models in our sample, $\mathcal{C}_{\rm
  max} = \mathcal{C}_{_{\rm TOV}}$ and we provide statistical evidence
that $\mathcal{C}_{\rm max} = 1/3$, in a way that is essentially
independent of the stellar mass.

%========================================================================
\section{Methods}
%========================================================================

An essential methodological part of our study is represented by the
construction of a large set of parameterised EOSs that provide a rich
ensemble through which statistical properties and general bounds can be
set. The construction of this set of EOSs is made in terms of the
parameterisation of the speed of sound and has been presented
by~\cite{Altiparmak:2022}. A number of different applications have been
made of this ensemble of EOSs~\cite{Ecker:2022, Ecker:2022b,
  Musolino2023b, Magnall2025}, underlying its validity and versatility.

In practice, our EOSs can be seen as the combination of different parts
whose constraints depend on the rest-mass densities considered. More
specifically, at the lowest densities, i.e., $n/n_s < 0.5$, -- where $n$
and $n_s := 0.16\,{\rm fm}^{-3}$ are the baryon number density and the
nuclear saturation density, respectively -- we use the
Baym-Pethick-Sutherland prescription~\cite{Baym71} for the crust. In the
range $0.5 \le n/n_s < 1.1$, we randomly sample polytropes to span the
range between the softest and stiffest EOSs
from~\cite{Hebeler:2013nza}\footnote{{It is in principle possible to use
  results from chiral effective field theory to extend to $n/n_s \simeq
  2$ the constraints for the low-density part of the EOSs~(see, \eg
  \cite{Drischler2020b}), albeit with singificatly larger relative
  errors. Although we find no change in our results when considering the
  subset of EOSs that are compatible with such
  constraints~\cite{Koehn2024}, we here prefer to employ a more
  conservative upper limit of $n/n_s=1.1$ because of the smaller relative
  errors.}}. At high densities ($n/n_s \approx 40$), corresponding to a
baryon chemical potential of $\mu = 2.6\,\rm GeV$, we impose the pQCD
constraint from~\cite{Fraga2014} on the pressure $p(X, \mu)$ of cold
quark matter, where the renormalization scale parameter $X$ is sampled
uniformly in the range $[1,4]$\footnote{{This constraint is by
  construction compatible with the integral constraint recently proposed
  by~\cite{Komoltsev:2021}.}}. {To assess the impact of the pQCD
  constraint on $\mathcal{C}_{\rm max}$, we also construct a separate
  ensemble in which this constraint is not imposed and finding
  considerable differences (see below).} For the intermediate density
range ($1.1\,n_s < n \lesssim 40\,n_s$), we follow~\cite{Annala2019} and
model the sound speed as piecewise-linear segments of the chemical
potential
\begin{equation}
  \label{eq:cs2}
  c_s^2(\mu) = \frac{\left(\mu_{i+1}-\mu \right) c_{s,i}^2 + \left(\mu -
    \mu_i \right) c_{s,i+1}^2}{\mu_{i+1}-\mu_i}\,,
\end{equation}
where $\mu_i$ and $c_{s,i}^2$ are parameters defining the $i$-th segment
in the range $\mu_i \leq \mu \leq \mu_{i+1}$. The number density is then
computed as
\begin{equation}
  \label{eq:n}
  n(\mu) = n_1 \exp \left({\int_{\mu_1}^\mu \frac{d\mu^\prime}{\mu^\prime
      c_s^2(\mu^\prime)}}\right)\,,
\end{equation}
where $n_1 = 1.1\,n_s$, and $\mu_1 = \mu(n_1)$ is set by the
corresponding polytropic EOS. The (isotropic) pressure\footnote{\cor{We
  note that we are not considering here anisotropic-pressure
  contributions as these would very naturally lead to compactnesses that
  are large and even compatible with those of black holes (see, \eg
  \cite{Raposo2018, Jampolski2024}).}} is obtained via
\begin{equation}
  \label{eq:p}
  p(\mu) = p_1 + \int_{\mu_1}^\mu d\mu^\prime \, n(\mu^\prime)\,,
\end{equation}
where the integration constant $p_1$ matches the pressure of the
polytrope at $n = n_1$, and we integrate Eq.~\eqref{eq:p} numerically
using seven segments for $c_s^2(\mu)$~(see \cite{Altiparmak:2022} for a
discussion).

Using this framework, we generate approximately $3 \times 10^5$ EOSs by
randomly sampling the free parameters $\mu_i \in [\mu_1, \mu_{N+1}]$
(where $\mu_{N+1} = 2.6\,\rm GeV$) and {$c_{s,i}^2 \in [0, c_{s,{\rm max}}^2]$ 
with uniformly distributed maximal sound speed $c_{s,{\rm max}}^2 \in [0, 1]$
to mitigate the undersampling of low sound-speed values.}
These EOSs are, by construction, consistent with nuclear theory and pQCD
uncertainties, and sufficiently numerous to reach the statistical
significance needed for our analysis. Also, while we do not explicitly
introduce strong first-order phase transitions for which $c^2_s=0$, EOSs
closely approximating first-order phase transitions, \ie with $c^2_s \sim
0.01-0.1$, are naturally present in our ensemble as a result of the
uniform sampling in the sound speed. {At the same time, when
  including the extensive coverage of suitably chosen EOSs studied
  by~\cite{Tan2022}, we find that our results apply unchanged also in
  the presence of strong first-order phase transitions.}

\begin{figure*}
 \centering     
 \includegraphics[width=0.495\textwidth]{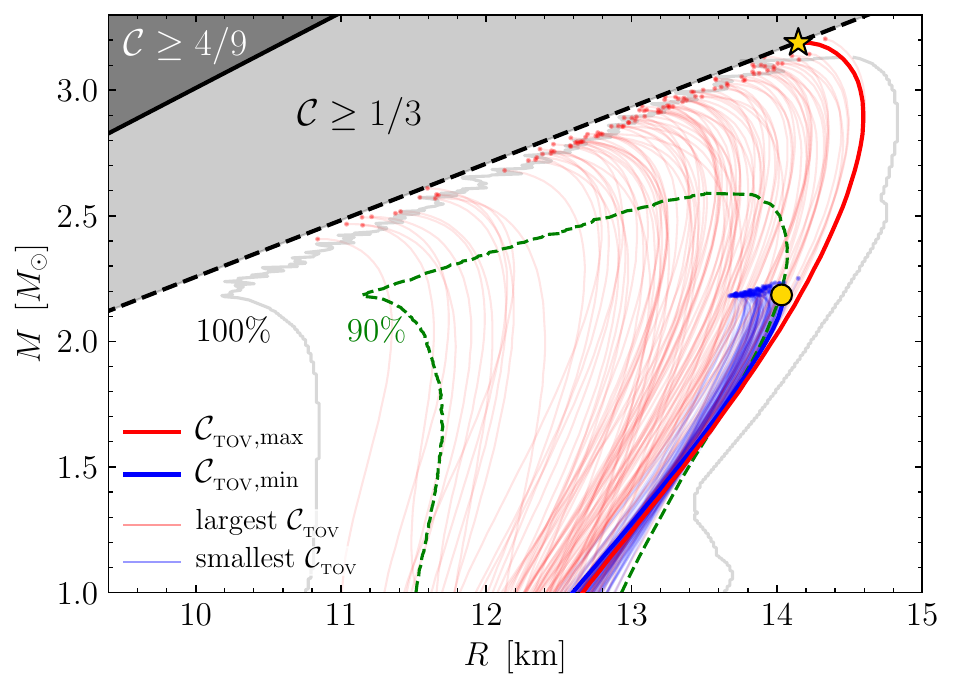}
 \includegraphics[width=0.495\textwidth]{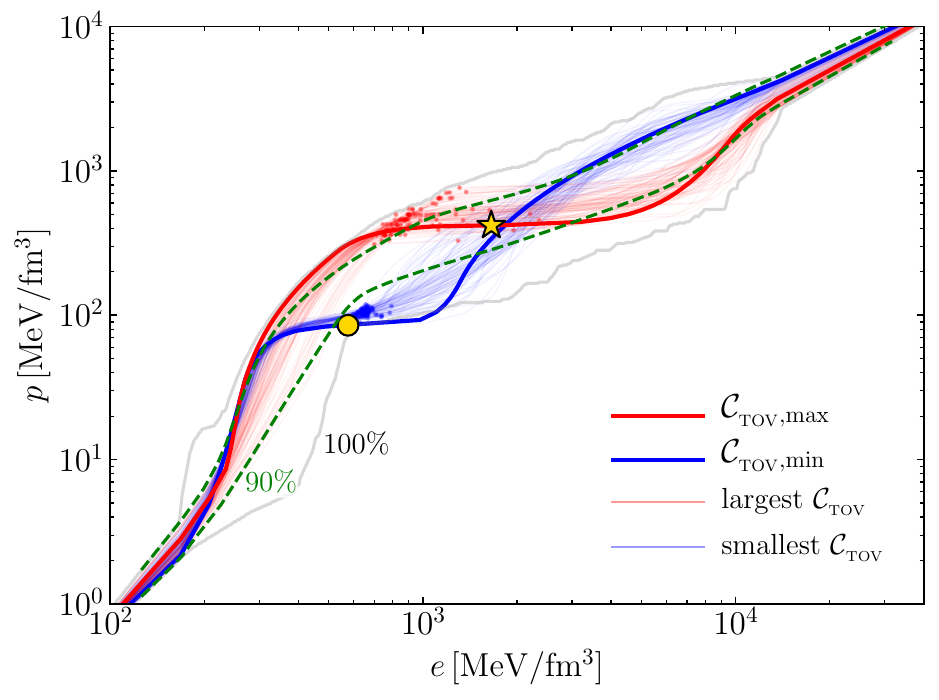}
 \caption{\textit{Left panel:} behaviour in the $(M,R)$ space of the
   sequences of nonrotating stars with EOSs leading to maximum-mass stars
   with the largest (smallest) compactness $\mathcal{C}_{_{\rm TOV}, {\rm
       max}}$ ($\mathcal{C}_{_{\rm TOV}, {\rm min}}$). Thin red (blue)
   lines refer to EOSs near the maximum (minimum) compactness, which is
   instead indicated with a thick red (blue) line. A golden star (circle)
   marks the position in the $(M,R)$ space of the star with the maximum
   (minimum) compactness of $\mathcal{C}_{\rm max} \simeq 0.3329$
   ($\mathcal{C}_{\rm min} \simeq 0.2294$) in the whole ensemble. Also
   shown with a solid grey (green dashed) line are the contours of the
   allowed ranges when considering $100\%$ ($90\%$) of the stars in the
   ensemble. Finally, shown with dark (light) grey shaded areas are the
   regions where $\mathcal{C} \geq 4/9$ ($\mathcal{C} \geq 1/3$).
   \textit{Right panel:} the same as on the left but in the $(p,e)$
   space.}
  \label{fig:EOSs} 
\end{figure*} 

For any EOS in our sample, we can then construct a sequence of
nonrotating stellar equilibria by solving the coupled set of TOV
equations
\begin{equation}
  \label{eq:TOV}
  \frac{dp(r)}{dr}
  =-\frac{\left[p(r)+e(r)\right]\left[\mathcal{C}(r)+4\pi r^2
      p(r)\right]}{r\left[1-2\mathcal{C}(r)\right]} = c_s^2\frac{de(r)}{dr}\,,
\end{equation}
\begin{equation}
  \label{eq:Mr}
  m(r)  = 4\pi \int_0^r e(r')\, r'^2 \, dr'\,,
\end{equation}
where $m(r)$ is the gravitational mass within the two-sphere of radius
$r$ and $\mathcal{C}(r) := m(r)/r$ with $r \leq R$. The last equality in
Eq.~\eqref{eq:TOV} employs the definition of the adiabatic sound speed
$c_s^2 := \left({dp}/{de}\right)_s$. In this way, we are able to
construct $2\times 10^8$ nonrotating stellar models in equilibrium, but,
of course, not all of these stellar configurations satisfy the present
astrophysical constraints. The latter can be expressed in terms of the
the mass measurements of J0348+0432~($M = 2.01\pm
0.04\,M_{\odot}$;~\cite{Antoniadis2013}) and J0740+6620~($M = 2.08 \pm
0.07\,M_{\odot}$;~\cite{Cromartie2019, Fonseca2021}) and the black-widow
binary pulsar PSR J0952-0607~($M=2.35\pm
0.17~M_\odot$;~\cite{Romani:2022jhd}), which we impose by discarding
EOSs yielding a maximum mass $M_{_{\rm TOV}} < 2.18\,M_{\odot}$. In
addition, we impose the NICER radius constraints from
J0740+6620~\cite{Miller2021, Riley2021} and J0030+0451~\cite{Riley2019,
  MCMiller2019b}, rejecting EOSs with $R < 10.75\,\rm km$ at $M =
2.0\,M_{\odot}$ or $R < 10.8\,\rm km$ at $M = 1.1\,M_{\odot}$. Finally,
we impose an upper bound on the binary tidal deformability
$\tilde{\Lambda}$ as deduced from GW170817 by rejecting all EOSs with
$\tilde{\Lambda} > 720$ (low-spin prior;~\cite{Abbott2018a}) at a chirp
mass $\mathcal{M}_{\rm chirp} = 1.186~M_\odot$ for mass ratios
$q>0.73$~(see also \cite{Magnall2025}, where this bound was not imposed
in the prior).

We should note that the loss of information following from discarding the
uncertainties associated with these measurements is effectively very
small, as discussed by~\cite{Jiang2022}, who showed that the underlying
distributions are almost identical regardless of the choice of constant
or variable likelihood, with $90\%$ credible intervals essentially
overlapping. Overall, as a result of this additional filtering process,
we construct $\approx 1.6\times 10^7$ stellar models satisfying all the
presently known theoretical and astrophysical constraints. These models
represent the basis of our statistical analysis.

%========================================================================
\section{Largest and smallest compactnesses}
%========================================================================

Using the methodology described above, it is possible to construct
probability density functions (PDFs) of the ensemble of stellar models in
the relevant space of parameters. The most interesting -- and commonly
employed -- ones are the space of masses and radii, and the space of
pressures and energy densities. These PDFs are shown respectively in the
left and right panels of Fig.~\ref{fig:EOSs}. More specifically, for each
panel, we show with a solid grey (green dashed) line the contours of the
allowed ranges when considering $100\%$ ($90\%$) of the stellar models in
the ensemble.

The $100\%$ confidence limits of the distributions in these two spaces
have been presented in a number of related works. Here we rather
concentrate on those EOSs that lead to the largest and smallest values of
the compactness measured for the maximum-mass star, \ie
$\mathcal{C}_{_{\rm TOV}, {\rm max}}$ and $\mathcal{C}_{_{\rm TOV}, {\rm
    min}}$, respectively. These are shown with solid thick red and blue
lines, while solid thin red and blue lines are used to report the 100
EOSs that have compactnesses close to the largest and smallest ones. Note
that each of these lines ends at the maximum-mass star and hence the gold
star (circle) marks the position of the stars with the largest (smallest)
compactness, while the thin little circles show the corresponding values
for the maximum-mass stars that have compactnesses close to the
largest/smallest one.

What can be easily appreciated from the left panel of Fig.~\ref{fig:EOSs}
is that stars with compactnesses near the maximum one actually span very
large ranges in radii and masses (thin red lines), with $11\,{\rm km}
\lesssim R \lesssim 14\,{\rm km}$, and $2.3\,M_{\odot} \lesssim M
\lesssim 3.3\,M_{\odot}$.  This behaviour follows from the degeneracy of
$\mathcal{C}$. Since the radius of high-mass neutron stars ($M >
2.20\,M_\odot$) can vary significantly, many different mass-radius
combinations yield nearly the same compactness. In our analysis, the
maximum compactness is attained by all those EOSs whose $M$–$R$ sequence
terminates in the central part of the upper edge of the grey $100\%$
interval shown in the left panel of Fig.~\ref{fig:EOSs}. By contrast,
the spread of the stellar models near the minimum compactnesses (thin
blue lines) is very small and the variance in radii is $\lesssim
0.4\,{\rm km}$, while that in the mass is $\lesssim 0.1\,M_{\odot}$. The
minimum compactness is clearly determined by the lower mass bound ($M >
2.18\,M_\odot$) imposed in our analysis, and its value corresponds to the
maximum allowed radius at this minimally permitted mass. We also note
that if strong first-order phase transition were to be present, then the
maximum-mass stars having the smallest-compactness would probably be
found at the onset of the phase transition.

To appreciate the origin of the different behaviour for EOSs near
$\mathcal{C}_{_{\rm TOV}, {\rm max}}$ and $\mathcal{C}_{_{\rm TOV}, {\rm
    min}}$, it is possible to look at the very distinct behaviour of the
EOSs associated with stars having maximum/minimum compactnesses. This is
shown in the right panel of Fig.~\ref{fig:EOSs} using the same convention
as in the left panel. Clearly, the maximum-compactness stars (thick and
thin red lines) correspond to rather stiff EOSs and the behaviour of the
pressure vs energy density essentially borders the $100\%$ confidence
contour.  Such EOSs typically lead to very massive stars with large-radii
and a fraction of them feature structures resembling first-order phase
transitions and are responsible for stars being not very massive and with
small radii.

On the other hand, the minimum-compactness stars (thick and thin blue
lines) correspond to the stiffest EOSs and experience a first-order phase
transition (or a rapid cross-over) at comparatively small energy
densities. The phase transition leads to an overall softening at large
densities and thus to sequences having smaller maximum masses. Naturally,
EOSs yielding stars with such large radii but small maximum masses are
also characterised by the smallest compactnesses.

Minimally compact stars are clearly realised by EOSs that are stiff at
low densities -- leading to large neutron star radii -- followed by an
extended region of low sound speed (i.e., small
pressure-vs-energy-density slope), beginning at $e_{_{\rm PT}} \approx
300\,\mathrm{MeV}/\mathrm{fm}^3$. As indicated by the blue small circles
lines, the corresponding maximum central densities of stable neutron
stars lie within the flat region around $e_{_{\rm TOV}} \approx
600\,\mathrm{MeV}/\mathrm{fm}^3$, signalling the termination of their
$M$–$R$ sequence. Conversely, maximally compact EOSs exhibit varying
stiffness at lower densities, leading to significant variance in neutron
star radii. They feature phase-transition onset densities that are rather
large and around $\epsilon_{_{\rm PT}} \approx
1000\,\mathrm{MeV}/\mathrm{fm}^3$, associated with a phase transition,
which typically coincides with $e_{_{\rm TOV}}$, indicating also in this
case that the transition marks the end of their $M$–$R$ sequence.

%========================================================================
\section{The most massive is the most compact}
%========================================================================

Having outlined our methodology, we next advance the conjecture that,
once an EOS is fixed, the star with the maximum mass also has the maximum
compactness $\mathcal{C}_{\rm max}$, that is,
\begin{equation}
  \label{eq:conjecture_1}
  \mathcal{C}_{\rm max}=\mathcal{C}_{_{\rm TOV}}\,.
\end{equation}
To substantiate this conjecture we use the fact that it is implied by the
stronger statement that the compactness is a monotonically growing
function of the gravitational mass
\begin{equation}
  \label{eq:conjecture_2}
  \frac{d\mathcal{C}(M)}{dM}
  = \frac{1}{R(M)}\left(1-\frac{d\ln R}{d\ln M}\right)
  \geq 0\,.
\end{equation}

Although the conjecture~\eqref{eq:conjecture_1} appears very natural and
almost intuitive, a rigorous analytical proof is lacking, at least to the
best of our knowledge. The root of the problem is in assessing whether
the term $d\ln R/d\ln M$ on the right-hand side of
Eq.~\eqref{eq:conjecture_2} is larger or smaller than unity. In turn,
this requires the integration of Eq.~\eqref{eq:Mr}, which would provide a
relation $M=M(R)$ (or, equivalently $R=R(M)$) for a generic EOS. Even
when making the additional assumption that the energy density is 
monotonically decreasing with $r$ inside the star, its nonlinear
behaviour prevents one from deriving a mathematical proof that $d\ln
R/d\ln M < 1$. This difficulty can also be cast in geometric terms:
assessing the size of $d\ln R/d\ln M$ amounts to measuring the slope of
the growth of the gravitational mass as a function of radius $M=M(R)$ in
a $(M,R)$ space~(see also~\cite{Ferreira2024} for a study of the slope in
agnostic EOSs). As shown in the left panel of Fig.~\eqref{fig:EOSs}, this
slope can vary considerably from EOS to EOS and even change sign when an
increase in the mass leads to stellar models with smaller radii.

That said, some progress can be made either considering specific analytic
solutions or by restricting the mathematical proof to a general class of
possible solutions (see End Matter). We start from the former and
consider a star with constant energy density $e_c$, or ``Schwarzschild
star''. In this case, the mass is simply given by $M=(4\pi/3)\,e_c\,R^3$,
so that $d\ln R/d\ln M = 1/3$, and thus $d\mathcal{C}/dM = (2/3)\,R >
0$. The second simple analytic example in support of the conjecture is
offered by the Tolman-VII (T-VII) solution~\cite{Tolman39}, that is often
invoked as convenient and analytic example of a compact star with
properties that are not too far from realistic neutron stars~(see, \eg
\cite{Lattimer01}). We recall that the energy density in a
``generalised'' T-VII solution is given by~\cite{Raghoonundun2015} $e(r)
= e_c (1 - \alpha {r^2}/{R^2})$, where $0 \leq \alpha \leq 1$ is a
constant introduced to modify the ``self-boundness'' and generalise the
original T-VII solution. Clearly $\alpha=1$ corresponds to the original
T-VII solution and $\alpha=0$ provides the Schwarzschild
star\footnote{Note that the energy-density profile of the generalised
T-VII solution is not necessarily zero at the stellar surface and,
indeed, $e(R)=e_c(1-\alpha)$, being zero only for $\alpha=1$, \ie the
original T-VII solution.}. A bit of algebra then shows that the relation
between the mass and the radius is given by
$M=(4\pi/3)(1-\alpha/5)e_c\,R^3$, so that, again, $d\ln R/d\ln M = 1/3$,
and thus $d\mathcal{C}/dM = (2/3)\,R > 0$, independent of the value of
$\alpha$.

\begin{figure}
  \centering     
  \includegraphics[width=0.66\textwidth]{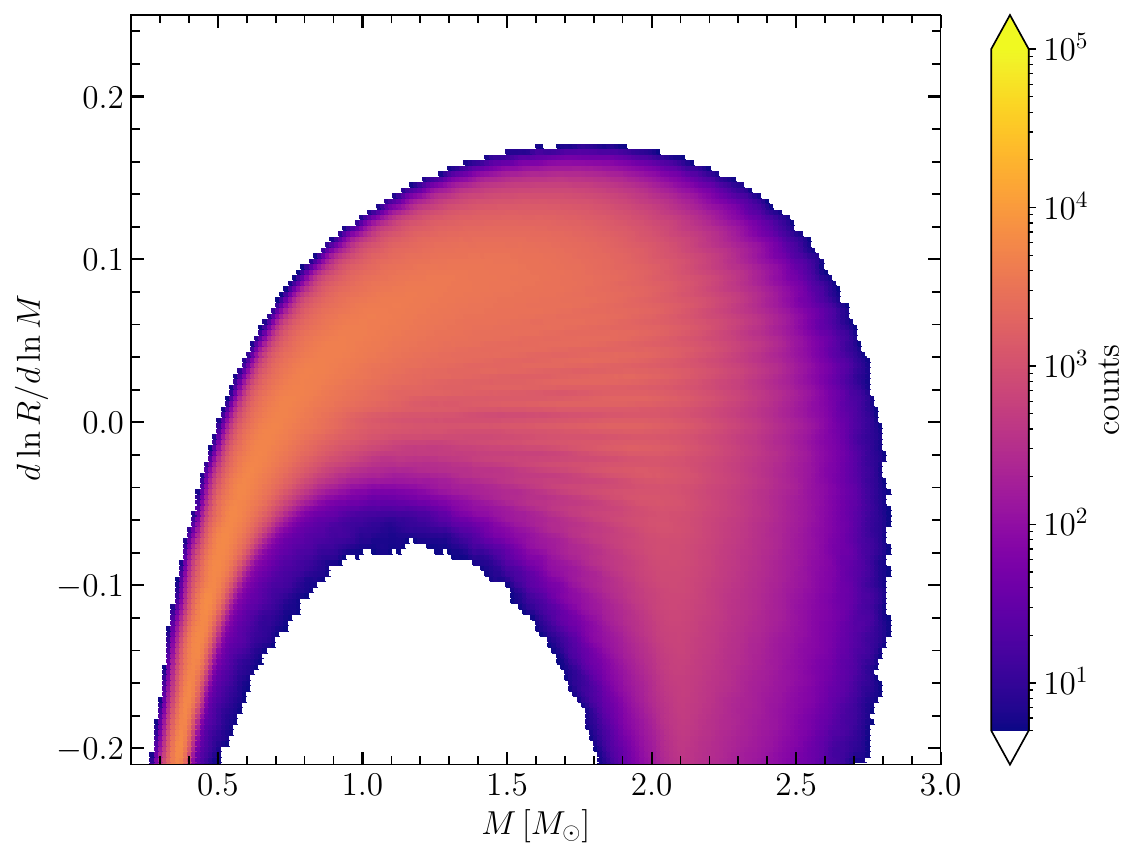}
 \caption{Distribution of the logarithmic derivative $d\ln R/d\ln M$ for
    all of the stellar models in our ensemble shown as a function of the
    stellar mass. Since $d\ln R/d\ln M \lesssim 0.18$, for each EOS the
    most massive star is also the most compact one.}
  \label{fig:dlnR}
\end{figure} 

Although the generalised T-VII solution offers a very good approximation
to a realistic neutron star, it does not exhaust all of the possible
behaviours of the energy density $e=e(r)$ and hence of the function
$M=M(R)$ for an arbitrary but consistent EOS. As a result, unable to
prove the validity of the conjecture~\eqref{eq:conjecture_1} in general,
we verify it by simply computing the logarithmic slope $d\ln R/d\ln M$
for all of the stellar models constituting our ensemble. This is shown in
Fig.~\ref{fig:dlnR}, which reports with colormap the distribution of the
logarithmic derivative $d\ln R/d\ln M$ as a function of the stellar
mass. Clearly, since ${d\ln R}/{d\ln M} \lesssim 0.18$ across all the
relevant range of masses, the function $\mathcal{C}(M)$ is monotonically
increasing with mass and the conjecture $\mathcal{C}_{\rm
  max}=\mathcal{C}_{_{\rm TOV}}$ is satisfied by all the stellar models
considered here. While this result does not have the rigour of a
mathematical proof, it does provide very strong evidence that, for
neutron-star models constructed with EOSs satisfying all known physical
and astrophysical constraints, the most massive stars are also the most
compact ones.

\begin{figure}
  \centering
  \includegraphics[width=0.66\textwidth]{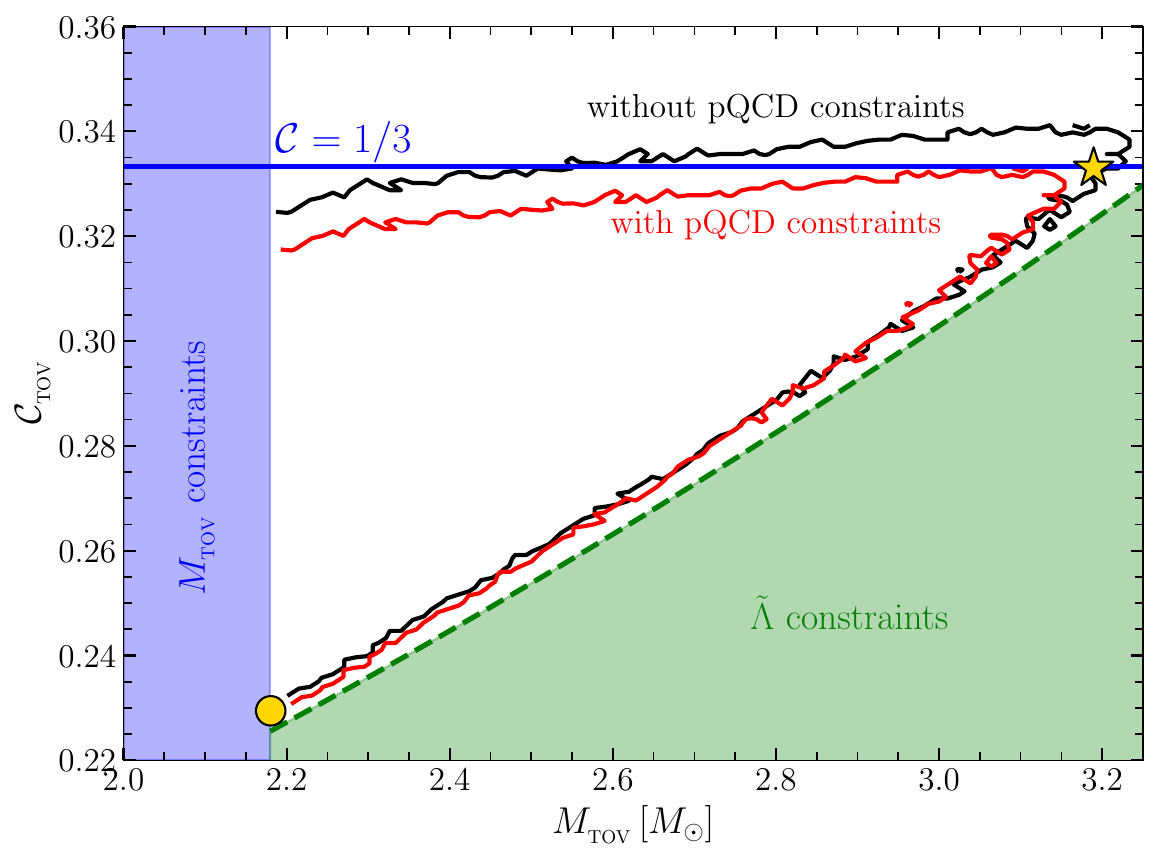}
  \caption{Distribution of the maximum compactness $\mathcal{C}_{_{\rm
        TOV}}$ as a function the maximum mass $M_{_{\rm TOV}}$. Shown as
    blue- and red-shaded areas are the constraints coming from the
    maximum mass and the binary tidal deformability, respectively. Also
    reported with a red dashed line is the analytic fit~\eqref{eq:Cmin}
    for the minimum of the maximum-mass compactness $\mathcal{C}_{_{\rm
        TOV}, {\rm min}}$, while the star and circle are the same models
    as in Fig.~\ref{fig:EOSs}. The outer bounds without imposing the pQCD
    constraints is shown with the black solid contour, {while the
      horizontal red solid line marks $\mathcal{C}=1/3$, highlighting
      that all stellar models are below this limit when the pQCD
      constraint is imposed.}}
  \label{fig:CMtov}
\end{figure} 

%========================================================================
\section{An upper limit on the compactness}
%========================================================================

Having shown the validity of the conjecture~\eqref{eq:conjecture_1} about
the maximum compactness, it is now possible to assess whether a global
upper limit exists to $\mathcal{C}_{\rm max}$. To this scope, all is
needed is to explore the distribution of the maximum-mass compactness
$\mathcal{C}_{_{\rm TOV}}$ for the ensemble of stellar models built. This
is shown in Fig.~\ref{fig:CMtov}, that reports with a colormap the
distribution of the maximum-mass compactness as a function of the maximum
mass $M_{_{\rm TOV}}$.

A rapid inspection of Fig.~\ref{fig:CMtov} reveals that the compactness
of the maximum-mass stars -- and hence the maximum compactness for a
given EOS -- has a clear upper limit and that this is given by
$\mathcal{C}_{\rm max} \simeq 0.3329 < 1/3$. Importantly, this limit is
essentially independent of the stellar mass and hence applies equally to
stars with maximum masses ranging from $M_{_{\rm TOV}} \sim 2.2$ to
$M_{_{\rm TOV}}\sim 3.0\,M_{\odot}$. Also important is that this upper
limit is essentially determined by the pQCD constraints, which
effectively prevent to have stars with very large masses and
comparatively small radii. {Indeed, stars with $\mathcal{C} > 1/3$
  can be found when the pQCD constraint is not imposed and as indicated
  with the black solid contour in Fig.~\ref{fig:CMtov}.}

Reported instead with a blue-shaded area is the region of the
$(\mathcal{C}_{_{\rm TOV}}, M_{_{\rm TOV}})$ space that is constrained by
the choice of the lower limit for $M_{_{\rm TOV}}$, so that the
blue-shaded area moves to the right for larger values of the assumed
minimum value of $M_{_{\rm TOV}}$ ($2.18\,M_{\odot}$ here). Finally,
marked with a red-shaded area is the region constrained by the limits
deduced on the binary tidal deformability $\tilde{\Lambda}$, which is
here set by the GW170817 event. The upper bound of the red-shaded area,
which marks the lower limit in the maximum-mass compactness
$\mathcal{C}_{_{\rm TOV}, {\rm min}}$. This limit is well-captured by the
quadratic relation
\begin{equation}
  \label{eq:Cmin} 	
  \mathcal{C}_{_{\rm TOV, min}}= c_1 + c_2\,\bar{M}_{_{\rm TOV}} +
  c_3\,\bar{M}^2_{_{\rm TOV}}\,,
\end{equation} 
where $\bar{M}_{_{\rm TOV}}:=\bar{M}_{_{\rm TOV}}/M_{\odot}$, and the
fitting coefficients are given by $c_1 = 0.100$, $c_2 =0.031$, $c_3 =
0.012$. Note that expression~\eqref{eq:Cmin}, which is shown with a thick
red dashed line in Fig.~\ref{fig:CMtov}, improves a similar expression
computed by~\cite{Most2020c} making use of the quasi-universal relation
between the maximum-mass of rotating and nonrotating
configurations~\cite{Breu2016}. This lower bound is determined by
constraints on tidal deformability and the resulting upper limits on
neutron-star radii inferred from GW170817.

Overall, the results presented in Fig.~\ref{fig:CMtov} lead to the
conclusion that the maximum compactness of neutron-star models satisfying
all known physical and astrophysical constraints is given by
$\mathcal{C}_{\rm max}= 1/3$, with the latter value determined by the
pQCD constraint.

%========================================================================
\section{Conclusions}
%========================================================================

Despite the significant recent progress, the knowledge of the properties
of matter in neutron stars still suffers from large uncertainties.  As a
result, a number of different EOSs have been derived under a variety of
assumptions and techniques, and all of these are routinely adopted when
modelling the structure and dynamics of neutron stars. Given these
difficulties, and at least at zero temperatures, it is possible to
approach the problem of the EOS from a purely statistical point of view
by generating a large ensemble of physically plausible EOSs constructed
so as to satisfy all the known physical constraints. In this way, despite
the large uncertainties, a number of robust results can be drawn simply
on statistical grounds.

Within this framework, we have considered a very basic and yet unanswered
question: \textit{how compact can a neutron star be?} We have addressed
this question by populating a very large ensemble of nonrotating stellar
models that are constructed making use of EOSs that satisfy all known
physics constraints. In addition, the population of stellar models
produced from such EOSs is further refined by imposing constraints coming
from astronomical observations. However, on the way to address the
question above, we have been faced with a related and equally basic
question: \textit{given an EOS, which star is the most compact one?}
Surprisingly, a simple answer does not exist to the best of our knowledge
and thus we conjectured that, given an EOS, the most massive star along
the sequence of equilibria is also the most compact or, equivalently,
that $\mathcal{C}_{\rm max} = \mathcal{C}_{_{\rm TOV}}$. This rather
natural conjecture can be proven to be true for some analytic solutions
and also for a generic family of stellar models, but remains unproven in
general. Luckily, the availability of our large ensemble of stellar
models has allowed us to prove, at least numerically, that the conjecture
is indeed true across our ensemble. The importance of the conjecture is
that it has allowed us to restrict our attention on the compactness of
the maximum-mass stars, which represent a much smaller set of stellar
models to consider. In this way, we have realised that the maximum
compactness for a given EOS has a clear upper limit that is almost
independent of the mass considered and given by $\mathcal{C}_{\rm max} <
1/3$. Because this bound does not appear when the pQCD constraints are
not imposed, it represents an intriguing imprint of pQCD at neutron-star
densities \cor{and of the softening that it induces at the highest
  densities and pressures~\cite{Altiparmak:2022, Gorda:2022}. In turn,
  the (unlikely) observation of a neutron star with compactness violating
  this bound would represent a terrific opportunity to reconsider
  critically the basic assumptions made in its derivation.}

\section*{Acknowledgements}
It is a pleasure to thank C. Providencia, K. Kokkotas,
J. Schaffner-Bielich, and A. Sedrakian for useful discussions. We are
also grateful to V. Dexheimer and V. Noronha-Hostler for kindly sharing
the set of EOSs published by~\cite{Tan2022}.

%% % TODO: include author contributions
%% \paragraph{Author contributions}
%% This is optional. If desired, contributions should be succinctly
%% described in a single short paragraph, using author initials.

% TODO: include funding information
\paragraph{Funding information}
Partial funding comes from the ERC Advanced Grant ``JETSET: Launching,
propagation and emission of relativistic jets from binary mergers and
across mass scales'' (Grant No. 884631). CE acknowledges support by the
Deutsche Forschungsgemeinschaft (DFG, German Research Foundation) through
the CRC-TR 211 ``Strong-interaction matter under extreme conditions''--
project number 315477589 -- TRR 211. LR acknowledges the Walter Greiner
Gesellschaft zur F\"orderung der physikalischen Grundlagenforschung
e.V. through the Carl W. Fueck Laureatus Chair.
%% Authors are required to provide funding information, including
%% relevant agencies and grant numbers with linked author's
%% initials. Correctly-provided data will be linked to funders listed in
%% the \href{https://www.crossref.org/services/funder-registry/}{\sf
%% Fundref registry}.

\begin{appendix}

\numberwithin{equation}{section}
  
%========================================================================
\section{A simple but partial proof of the conjecture}
%========================================================================

In the main text we have discussed how it is hard to prove analytically
the validity of the conjecture~\eqref{eq:conjecture_1}, although we do
not exclude that a proof is possible. However, we do provide here the
mathematical details that allow to prove the validity of the
conjecture~\eqref{eq:conjecture_1} for a specific but sufficiently
general class of stellar models. In particular, we will consider a broken
power law where $dM/dR$ changes sign at a given mass $M_*$ and radius
$R_*$, much like the behaviour shown by most of the EOSs in the left
panel of Fig.~\ref{fig:EOSs}. For convenience, we cast this result in
terms of the following theorem.\\ \\
\textbf{\textit{Theorem.~}} Given a perfect fluid described by a
barotropic equation of state, the solutions of the Einstein equations
leading to static and spherically symmetric equilibrium configurations
with mass $M$, radius $R$, and compactness $\mathcal{C}:=M/R$, are such
that the stellar model with the maximum mass $M_{_{\rm TOV}}$ is also the
most compact one when
\begin{equation}
  \label{eq:scaling}
  M(R) = \left\{
  \begin{array}{lcccc}
    \kappa_1 \, R^p   & \quad {\rm for}& \quad &M \leq M_*& \\
    \kappa_2 \, R^{-q} & \quad {\rm for}& \quad &M \geq M_*& \,,
  \end{array}
  \right.
\end{equation}
where $\kappa_1, \kappa_2, p$ and $q$ are real and positive
constants. Imposing continuity at $M_*$ then constraints $\kappa_2 =
\kappa_1 R^{p+q}_*$, where $R_*$ is the radius where $M(R_*)=M_*$.  \\

\textit{Proof.~} We proceed along the same logical route anticipated in
the main text, that is, by proving that the compactness is a
monotonically growing function such that $\mathcal{C}_{\rm
  max}=\mathcal{C}_{_{\rm TOV}}$. In turn, this implies proving that [see
  Eq.~\eqref{eq:conjecture_2}]
\begin{equation}
  \label{eq:logder}
  \frac{d\ln R}{d\ln M} \leq 1\,.
\end{equation}

Using the proposed general scaling~\eqref{eq:scaling}, it is the simple
to compute that 
\begin{equation}
  \label{eq:scaling_2}
 \frac{d\ln R}{d\ln M} = \left\{
 \begin{array}{lcccc}
   \phantom{-}{1}/{p} & \quad &{\rm for}& \quad M \leq M_*& \\
             -{1}/{q} & \quad &{\rm for}& \quad M \geq M_*& \,,
  \end{array}
  \right.
\end{equation}
so that the condition~\eqref{eq:logder} [and hence the
  conjecture~\eqref{eq:conjecture_1}] is satisfied if
\begin{equation}
  \label{eq:n_condition}
  \left\{
  \begin{array}{lcccc}
   p \geq 1   & \quad {\rm for}& \quad &M \leq M_*& \\
   q \geq -1  & \quad {\rm for}& \quad &M \geq M_*& \,.
  \end{array}
  \right.
\end{equation}

A few comments are worth making, some of which can be seen as corollaries
of the theorem.
\begin{itemize}

\item the scaling~\eqref{eq:scaling} is only a sufficient condition for
  the validity of the conjecture~\eqref{eq:conjecture_1}.

\item the scaling~\eqref{eq:scaling} is of class $\mathscr{C}^0$ at $R_*$, \ie
  continuous but with discontinuous derivatives. If useful, additional
  scaling terms can be introduced to guarantee also continuity of the
  higher derivatives.

\item the generalised T-VII solution represents a special case of the
  class~\eqref{eq:scaling} for $M_* \to \infty$ and $p=3$. Using the
  condition~\eqref{eq:n_condition}, it is simple to deduce that the
  generalised T-VII solution trivially satisfies the
  conjecture~\eqref{eq:conjecture_1}.

\item a modification of the T-VII solution has also been proposed in
  which the behaviour of the energy density is extended by the
  introduction of a quartic term~\cite{Jiang2019}
  \begin{equation}
    \label{eq:mTVII}
    e(r) = e_c \left(1 -
    \tilde{\alpha}\,{r^2}/{R^2} - (\tilde{\alpha} -
  1){r^4}/{R^4}\right)\,.
  \end{equation}
This solution is also referred to as the ``modified'' T-VII solution and
$\tilde{\alpha} \in [0, 2]$ is a new parameter such that
$\tilde{\alpha}=1$ leads to the generalised T-VII solution and
$\tilde{\alpha}=0$ to the Schwarzschild solution~(see \cite{Posada2021}
for a discussion of the role played by $\tilde{\alpha}$). Although
expression~\eqref{eq:mTVII} may appear more complex, it is easy to show
that it effectively belongs to the class~\eqref{eq:scaling} with $M_* \to
\infty$ and $p=3$. Hence, also the modified T-VII solution trivially
satisfies the conjecture~\eqref{eq:conjecture_1}.
  
\item although the generalised T-VII solution does not have a maximum
  mass, the Buchdahl limit constraints the central energy density to be
\begin{equation}
e_c < \frac{1}{3\pi\, R^2 (1 - 3\alpha/5)} \,,
\end{equation}
so that, even in the absence of a maximum-mass star, it is still true
that the most massive star is also the most compact one. The same
conclusion is true for the modified T-VII solution, in which case the
limit on the central energy density reads
\begin{equation}
e_c < \frac{1}{3\pi\, R^2 \left[1 - 3\tilde{\alpha}/5 + 3(\tilde{\alpha}
    - 1)/7\right]} \,.
\end{equation}
\end{itemize}

%========================================================================
\section{More on the largest and smallest compactnesses}
%========================================================================

\begin{figure*}
  \centering     
  \includegraphics[width=0.495\textwidth]{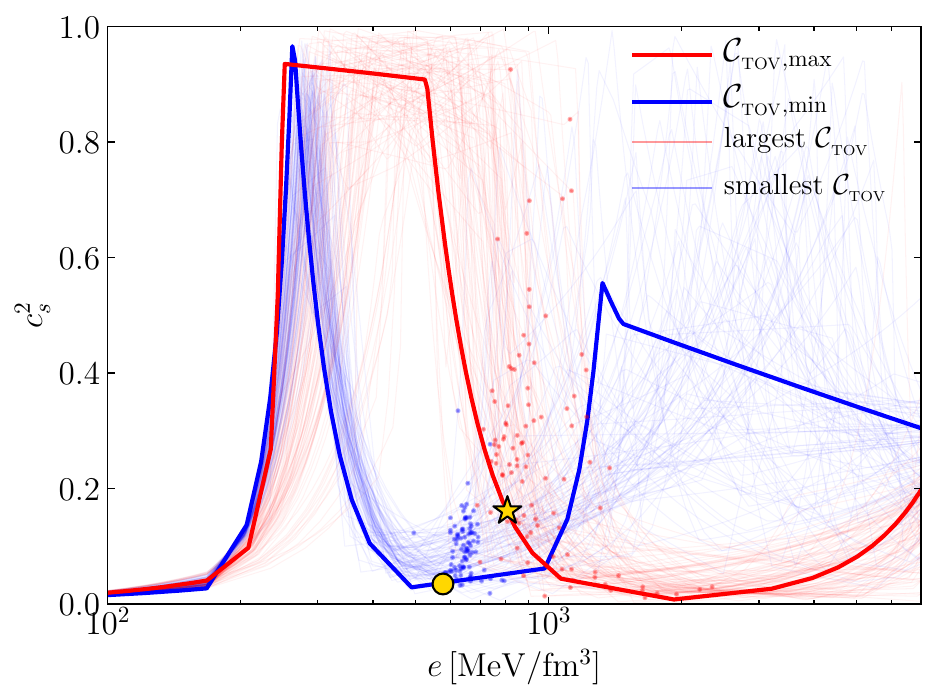}
  \includegraphics[width=0.495\textwidth]{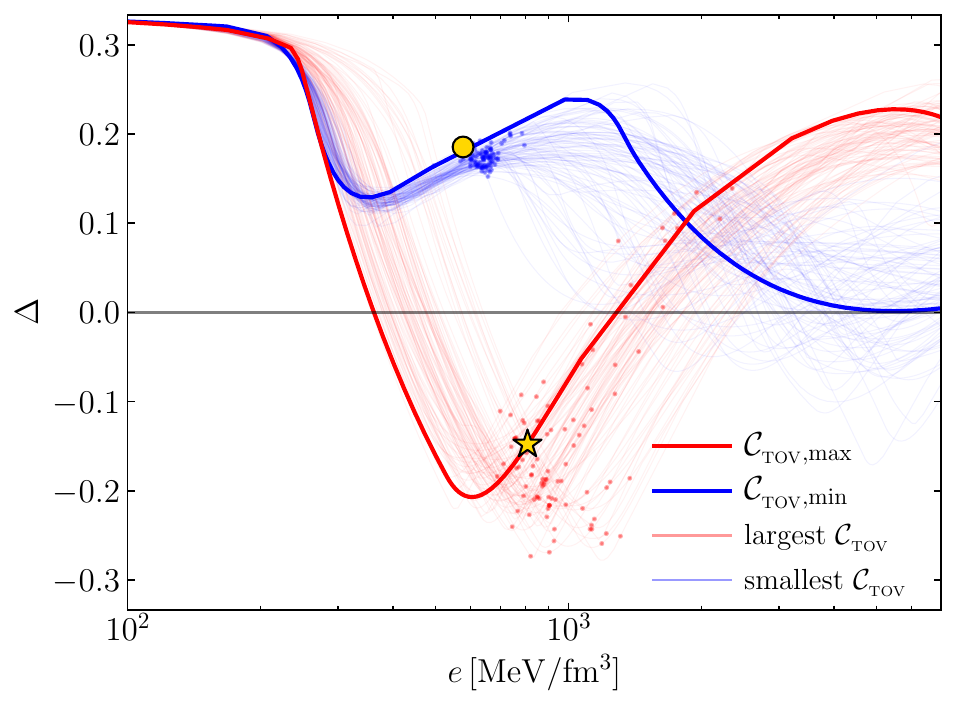}
  \caption{The same as in Fig.~\ref{fig:EOSs} but for the behaviour of
    the sound speed (left panel) and of the conformal anomaly as a
    function of the energy density (right panel).}
 \label{fig:EOSs_2} 
\end{figure*}

In the main text we have discussed some of the properties of the stellar
models belonging to the class of EOSs leading the largest and smallest
values of the maximum-mass compactnesses, \ie $\mathcal{C}_{_{\rm TOV},
  {\rm max}}$ and $\mathcal{C}_{_{\rm TOV}, {\rm min}}$. We have also
employed Fig.~\ref{fig:EOSs} to show the properties of such EOSs in terms
of their behaviour in the $(M,R)$ and $(p,e)$ spaces; we here use
Fig.~\ref{fig:EOSs_2} to provide some additional information. More
specifically, using the same colour and line-style conventions adopted in
Fig.~\ref{fig:EOSs}, the left panel of Fig.~\ref{fig:EOSs_2} reports the
behaviour of the square of the sound speed as a function of the energy
density, where it is very easy to realise that the EOSs leading to the
maximum compactness have a rapid cross-over -- as indicated by the sound
speed dropping to very small values -- only at very large energy
densities, thus leading to hybrid stars with a very small quark core. By
contrast, the EOSs leading to the minimum compactness have a first-order
phase transition already at very low energy densities (in both cases
$c^2_s \to 1/3$ for $n \gg n_s$).

The right panel of Fig.~\ref{fig:EOSs_2} shows instead the functional
behaviour of the conformal anomaly $\Delta := 1/3 - p/e$, where $p$ is
the pressure, and where $ -2/3 \leq \Delta \leq 1/3$ for thermodynamical
stability (we recall that $\Delta \to 0$ for $n \gg n_s$). Worth noting
in this case is that the conformal anomaly is always positive for EOSs
with compactnesses around $\mathcal{C}_{_{\rm TOV}, {\rm min}}$, while it
becomes negative in large portions of the star for EOSs with
compactnesses close to $\mathcal{C}_{_{\rm TOV}, {\rm max}}$; this is a
behaviour already noted in a number of papers (see, \eg
\cite{Fujimoto:2022, Marczenko:2022, Ecker:2022b}).

%% \section{First appendix}
%% Add material which is better left outside the main text in a series of Appendices labeled by capital letters.

%% \section{About references}

%% Your references should start with the comma-separated author list
%% (initials + last name), the publication title in italics, the journal
%% reference with volume in bold, start page number, publication year in
%% parenthesis, completed by the DOI link (linking must be implemented
%% before publication). If using BiBTeX, please use the style files
%% provided on
%% \url{https://scipost.org/submissions/author_guidelines}. If you are
%% using our LaTeX template, simply add

%% \begin{verbatim}
%% \bibliography{your_bibtex_file}
%% \end{verbatim}
%% at the end of your document. If you are not using our LaTeX template, please still use our bibstyle as
%% \begin{verbatim}
%% \bibliographystyle{SciPost_bibstyle}
%% \end{verbatim}
%% in order to simplify the production of your paper.
\end{appendix}

%%%%%%%%% END TODO: CONTENTS

%%%%%%%%%% TODO: BIBLIOGRAPHY
% Provide your bibliography here. You have two options:

%%% FIRST OPTION
% Write your entries here directly, following the example below, including:
% Author(s), Title, Journal Ref. with year in parentheses at the end, followed by the DOI number.

%\begin{thebibliography}{99}
%\bibitem{1931_Bethe_ZP_71} H. A. Bethe, {\it Zur Theorie der Metalle. i. Eigenwerte und Eigenfunktionen der linearen Atomkette}, Zeit. f{\"u}r Phys. {\bf 71}, 205 (1931), \doi{10.1007\%2FBF01341708}.
%\bibitem{arXiv:1108.2700} P. Ginsparg, {\it It was twenty years ago today... }, \url{http://arxiv.org/abs/1108.2700}.
%\end{thebibliography}

%%% SECOND OPTION
% Use your bibtex library, formatted by the SciPost style file.
\bibliography{aeireferences}
%\bibliographystyle{SciPost_bibstyle}

%% \bibliography{SciPost_Example_BiBTeX_File.bib}

%%%%%%%%%% END TODO: BIBLIOGRAPHY

\end{document}